%% file: 0main.tex
\documentclass[sigconf,natbib=true,anonymous=false]{acmart}
\usepackage{pifont}
\usepackage{url}
\usepackage{subcaption}
\usepackage{tabularx, booktabs}

\AtBeginDocument{%
  \providecommand\BibTeX{{%
    \normalfont B\kern-0.5em{\scshape i\kern-0.25em b}\kern-0.8em\TeX}}}

\setcopyright{acmcopyright}
\copyrightyear{2023}
\acmYear{2023}
\acmDOI{XXXXXXX.XXXXXXX}

\acmConference[SIGIR '23]{The 46th International ACM SIGIR Conference on Research and Development in Information Retrieval}{July 23--27, 2023}{Taipei}
\acmBooktitle{The 46th International ACM SIGIR Conference on Research and Development in Information Retrieval,
  July 23--27, 2023, Taipei}
\acmPrice{15.00}
\acmISBN{978-1-4503-XXXX-X/18/06}
\acmPrice{15.00}
\acmISBN{978-1-4503-XXXX-X/18/06}

\begin{document}

\title{Improving Legal Case Retrieval with Brain Signals}

\author{Ruizhe Zhang}
\affiliation{%
  \institution{Quan Cheng Laboratory, \\
Dept. of CS\&T, Institute for Internet Judiciary, Tsinghua University}
  \city{Beijing}
  \country{China}
  \postcode{100084}
}
\author{Qingyao Ai}
\affiliation{%
  \institution{Quan Cheng Laboratory, \\
Dept. of CS\&T, Institute for Internet Judiciary, Tsinghua University}
  \city{Beijing}
  \country{China}
  \postcode{100084}
}
\author{Ziyi Ye}
\affiliation{%
  \institution{Quan Cheng Laboratory, \\
Dept. of CS\&T, Institute for Internet Judiciary, Tsinghua University}
  \city{Beijing}
  \country{China}
  \postcode{100084}
}

\author{Yueyue Wu}
\affiliation{%
  \institution{Quan Cheng Laboratory, \\
Dept. of CS\&T, Institute for Internet Judiciary, Tsinghua University}
  \city{Beijing}
  \country{China}
  \postcode{100084}
}
\author{Xiaohui Xie}
\affiliation{%
  \institution{Quan Cheng Laboratory, \\
Dept. of CS\&T, Institute for Internet Judiciary, Tsinghua University}
  \city{Beijing}
  \country{China}
  \postcode{100084}
}
\author{Yiqun Liu \footnotemark[1]}
\affiliation{%
  \institution{Quan Cheng Laboratory, \\
Dept. of CS\&T, Institute for Internet Judiciary, Tsinghua University}
  \city{Beijing}
  \country{China}
  \postcode{100084}
}
\email{yiqunliu@tsinghua.edu.cn}


\begin{abstract}

The tasks of legal case retrieval have received growing attention from the IR community in the last decade.
Relevance feedback techniques with implicit user feedback (e.g., clicks) have been demonstrated to be effective in traditional search tasks (e.g., Web search)
In legal case retrieval, however, collecting relevance feedback faces a couple of challenges that are difficult to resolve under existing feedback paradigms.
First, legal case retrieval is a complex task as users often need to understand the relationship between two legal cases in detail to correctly judge their relevance. 
Traditional feedback signal such as clicks on SERPs is too coarse to use as they do not reflect any fine-grained relevance information.
Second, legal case documents are usually long, and users often need several or even tens of minutes to read and understand them. 
Simple behavior signal such as mouse clicks and eye-tracking fixations can hardly be useful when users almost click and examine every part of the document.
In this paper, we explore the possibility of solving the feedback problem in legal case retrieval with brain signal. Recent advances in brain signal processing have shown that human emotional information can be collected in fine grains through Brain-Machine Interfaces (BMI) without interrupting the users in their tasks. 
Therefore, we propose a framework for legal case retrieval that uses EEG signal (brain signal collected with a specific type of BMI) to optimize retrieval results. We collected and create a legal case retrieval dataset with users’ EEG signal and propose several methods to extract effective EEG features for relevance feedback. Our proposed features achieve a 71\% prediction accuracy for feedback prediction with an SVM-RFE model, and our proposed ranking method that takes into account the diverse needs of users can significantly improve user satisfaction for legal case retrieval. Experiment results show that re-rank result list makes the user more satisfied than the fixed result list.

\end{abstract}



\received{1 February 2023}
\received[revised]{1 March 2023}
\received[accepted]{1 January 2023}

\maketitle
\input{1introduction}
\input{2relatedWork}

\input{3_taskdefinition}

\input{5methods}
\input{6dataCollection}
\input{7examinations}
\input{8conclusion}
\begin{acks}

\end{acks}
\newpage
\bibliographystyle{ACM-Reference-Format}
\bibliography{sample-base}

\appendix

\end{document}

%% file: 1introduction.tex
\section{Introduction}
With the development of information retrieval technology, it has become a common practice for legal practitioners to use legal case retrieval systems to organize and search legal documents in their daily work. As the quality of such systems could significantly impact both the efficiency and accuracy of user's decision making in downstream tasks, how to construct legal case retrieval system has received considerable attention in recent years.  

Among different retrieval techniques, relevance feedback has been proven to be effective for almost all types of retrieval systems \cite{ruthven2003survey}. With implicit or explicit user feedback, search engines can improve retrieval performance by re-ranking or re-weighting information based on their relationships to the feedback documents. Particularly, relevance feedback collected from user behaviors~(e.g., clicks, hovers, eye fixations, etc.) have shown to be important for the optimization of Web search engines \cite{harman1992relevance}. Therefore, it's tempting to apply relevance feedback techniques to improve legal case retrieval.
However, there are several challenges that limit the effectiveness of relevance feedback in legal case retrieval. First, in contrast to open-domain retrieval~(e.g., Web search), legal case retrieval systems are designed for legal professionals who need to find and analyze different case judgements for their downstream legal tasks. Judgment is a legal document that is used to record the case and its verdict. In the legal case retrieval, users' search targets are judgments, instead of documents in web search. The relations between case judgements and user’s information needs tends to be complicated and, in many cases, fine-grained. To avoid interrupting user’s search experience, existing relevance feedback methods are usually built with implicit feedback signals such as clicks \cite{craswell2008experimental,Accuratelyinterpretingclickthroughdataasimplicitfeedback,inproceedings,chapelle2009dynamic,arguello2009sources},  and hovers \cite{o2016leveraging,huang2011no,buscher2012large}. Unfortunately, such signals are too coarse to be useful in legal case retrieval because they cannot capture user’s actual opinions on case judgements in fine granularity. 

Second, due to the complexity of downstream legal tasks and the extreme length of legal judgements (e.g., $8.2k$ words in average according to our statistics of LeCard~\cite{LeCard}), user behaviors in legal case retrieval often have significant different patterns with those in Web search. For example, because it’s difficult to judge the usefulness of a case judgements just from its title and abstract, legal case retrieval users tend to click and check multiple candidates carefully and thoroughly before making any decisions. This leads to significant more clicks and eye fixations on judgements no matter whether the search results are relevant or not~\cite{InvestigatingUserBehaviorinLegalCaseRetrievalSYQ}. Traditional relevance feedback methods could easily fail in such scenarios as they assume that ``more interactions means more relevant'' in search.

For the above reasons, the feedback signals must reflect user's logical thinking process and the usefulness of case judgements in fine granularity. Effective relevance feedback in legal case retrieval must have at least the following two characteristics. (1)~The feedback signals must reflect user's logical thinking process and the usefulness of case judgements in fine granularity. (2)~The relevance feedback process needs to conducted in a seamless way with minimum interruptions on user's search experience.


Inspired by recent advances on Brain-Machine Interface (BMI) technology, in this paper, we propose a BMI-based relevance feedback framework for legal case retrieval.
With the rapid development of non-injection BMI devices, collecting and analyzing brain signals is becoming less expensive and more feasible in practice. Recent studies have shown that, with small wearable devices, we can collect considerable meaningful brain signals without interpreting users’ searching and learning process \cite{nicolas2012brain,moshfeghi2016understanding,paisalnan2022neural}. As a special type of BMI devices, EEG devices and the brain signals they collected, i.e., EEG signals, have already been shown to be useful in analyzing user’s behavior and search satisfaction in open-domain retrieval tasks such as Web search~\cite{pinkosova2020cortical,gwizdka2017temporal}. We believe that such information could benefit the design of legal retrieval system and potentially improve the effectiveness of relevance feedback in legal case retrieval.
 
Specifically, our proposed framework include three steps. First, based on the user’s query, a initial legal case retrieval system is deployed to retrieve a small set of candidate cases from the judgement corpus. Second, we show the candidate results one by one and collect user’s EEG data with a wearable BMI device. After that, we construct a legal EEG module to analyze user’s feedback through their brain signals and use its output to re-rank future judgements to show. Particularly, existing EEG methods mostly focus on short and simple stimuli~(e.g., sounds, flashes, pictures) analysis, which cannot fit our needs due to the complexity and long-time duration of legal search tasks. To this end, we propose a new EEG analysis method by extracting and analyzing EEG signals at different frequency domains in different time periods. The proposed EEG module can extract effective features that boost the accuracy of feedback prediction to nearly 80\% with simple machine learning methods such as SVM-RFE. Our simulation experiments and lab study of 20 domain experts have shown that our EEG-based relevance feedback framework can significantly improve the performance of legal case retrieval.

The main contributions of this article include the following four components:
\begin{itemize}
    \item We proposed the first brain signal supported legal case retrieval framework.
    \item Within this framework, a set of user experiments was designed. We recruited 20 participants to participate in the experiment and collected their EEG datasets.
    \item We designed a method to analyze long-term EEG signals that can be embedded in the framework. 
    \item We have tested the method with the collected dataset and report the experimental results. Our method is efficiently for complex work such as legal case retrieval.
\end{itemize}


%% file: 2relatedWork.tex
\section{Related Work}
Since this paper is about how to improving legal case retrieval with brain signals, we will present these three important related work: BMI technology, legal case retrieval and relevance feedback.
\subsection{BMI}
Brain Machine Interface~(BMI) is an novel communications system that utilize brain signals to enhance the interactions between human and external devices~\cite{nicolas2012brain}.
Recently, BMI is becoming portable and low-cost~\footnote{https://the-unwinder.com/reviews/best-eeg-headset/} for real-world applications, such as game playing~\cite{van2012designing}, education~\cite{chen2022inter}, and so on.
In the domain of human-centered information retrieval~(IR), BMI was also widely applied to improve search experiences.
Beyond the basic usage of BMI in IR, i.e., controlling search interface~\cite{chen2022web}, BMI was also widely applied to help us understanding brain activities during search process.
For example, \citet{moshfeghi2016understanding,paisalnan2022neural} explored and analyzed brain signals during the arisen and satisfaction of information need.
~\citet{ye2022towards} observed the neurological difference between brain responses to different textual contents in a reading comprehension scenario.
The relevance judgment process during IR tasks was also widely studied with neurological methods~\cite{pinkosova2020cortical,gwizdka2017temporal} and it was suggested that internal links  between relevance and brain activities exist.

Although prior literature has explored using brain signals to monitor human's brain responses to textual contents~\cite{ye2022towards, pinkosova2020cortical, barral2017bci} and attempted to understanding their relevance judgments~\cite{pinkosova2020cortical,gwizdka2017temporal}, few research has explored the brain activities during legal related tasks.
A important difference between traditional IR and legal case retrieval is that the textual content in a legal case are usually longer than a common document in Web search and requires a long-term examinations.
Hence,  directly bringing in EEG techniques from traditional IR scenarios into legal scenarios are infeasible and the main challenge is the long-term modeling of brain signals.
To uncover this issue, we adopt a novel feature extraction for the application of BMI in legal related scenarios.

\subsection{Legal Case Retrieval}

From 1990s, more and more legal practitioners have been improving their work efficiency by legal case retrieval \cite{legalSearch1997}. 
The goal of the legal case retrieval is to find judgments that are similar to the current case in terms of merits or logic. 
In general, the judgement database contains a large number of judgements from cases in which judgements have been delivered. 
Users searching for judgments to assist them in making relevant judgments about current cases or justified her judgment in court.
In common law systems, after finding facts, a lawyer or judge needs to locate any relevant statutes and cases~\citeN{alma99158810270101841,enwiki:1131543007,garner2001dictionary}.
The lawyer must extract the principles, analogies and statements the similarities between two cases to convince the judge of the applicable law in current case.
Judges must also decide current cases appropriately based on analogous cases in common law systems.
In civil law systems, customary law systems or other law systems, legal practitioners are also using legal case search to determine the legal issues in the current case and to forecast possible judgments or make fair judgments \cite{rehberg1998accidental}.

In response, legal case retrieval has also received increasing attention from researchers \cite{shao2020towards,tran2020encoded,van2017concept}. Some models of web retrieval have been adapted to legal case retrieval, and some models specifically for legal case retrieval have also been proposed \cite{shao2020bert}.

In web search, a series of traditional algorithms are used for ranking, such as BM25 \cite{robertson1994some}, TF-IDF \cite{salton1988term} etc. Also, deep learning algorithms like LTR(learning to rank) \cite{liu2011learning}, DSSM \cite{hu2014convolutional}, RNN~\cite{pang2016text}, CNN \cite{shen2014latent}, SRNN \cite{wan2016match} have been applied to optimize the efficient of ranking. 

\subsection{Relevance Feedback}

In 1990, \citet{salton1990improving} introduce relevance feedback in information retrieval. Then, relevance feedback has become a hot problem in research. At first, researchers collected explicit relevance feedback to optimize retrieval results \cite{harman1992relevance,ruthven2003survey}. Later, researchers collected user click signals to determine how users rated the results \cite{craswell2008experimental,Accuratelyinterpretingclickthroughdataasimplicitfeedback,inproceedings,chapelle2009dynamic,arguello2009sources}. In the field of image search, hover can be collected by browsers and reflects user preference \cite{o2016leveraging,huang2011no,buscher2012large}. \citet{buscher2008eye} suggested that eye movement can also be used as implicit relevance feedback \cite{cutrell2007you,guan2007eye,buscher2010good,papoutsaki2017searchgazer}. 
Through this relevance feedback, the search engine is able to probe the user's preferences. Based on user preferences, search engines re-rank results to optimize users' satisfaction.

However, user behavior in legal case search is very different from web search, specifically in session length, target of search, user behavior etc \cite{InvestigatingUserBehaviorinLegalCaseRetrievalSYQ}.
In legal case search, the logically complex results make it difficult for users to distinguish whether a result is useful or not in a short period of time. Traditional feedback signals become difficult to exploit \cite{yuan1997end}.

%% file: 3_taskdefinition.tex
\section{task definition}
The scenario of legal case retrieval is different from the web search scenario. In order to introduce the scenarios of legal case retrieval more clearly, we describe the contents and purpose of legal case retrieval users in detail here.

\begin{figure}[ht]
    \flushleft
    \includegraphics[scale=0.3]{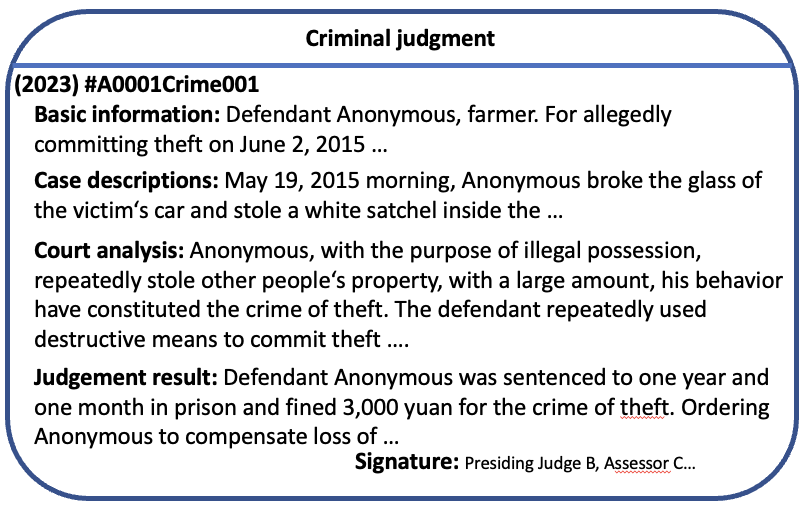}
    \caption{An example legal case in our corpus. A legal case composes of four parts: \textit{Basic information, Case description, Court analysis} and \textit{Judgement result}.}
    \label{judgement}
\end{figure}

Legal case retrieval is a frequent part of legal practitioners' work.
The core of the legal case search is the judgments. A judgment is structured as shown in Figure \ref{judgement}. This structure is fixed and has a strong reminder to the reader. In the publicly available judgments database, all judgments are structured according to the structure as in Figure \ref{judgement}.

In this paper, we focus on the legal case retrieval of criminal field as our research targets. 
In the field of criminality, both civil law and common law systems follow the principle of "No crime and no penalty without a previous penal law" (also known as "the legality principle").
In practice, every criminal indictment and criminal judgment will involve one or more charges. 
In this work, we consider the key factor of the judgment, "charge(s) involved", as the subtopic(s) of this judgment for the purpose of analysis and discussion.

Generally, professional users use the legal case retrieval engine mainly for the purpose of retrieving judgments of other cases which are similar to the case they need to deal with.
Before starting their search, professional legal users will organize a case description of the case they need to deal with and write it down as a complete text. That is to be clear, "What the fact is". 
In legal cases, text that clearly describes the facts of the case is often long. The length of such texts is often much longer than query words in web searches \cite{InvestigatingUserBehaviorinLegalCaseRetrievalSYQ}. In this experiment, we use the "description" part of the LeCard dataset as the user's query term. That is, the user enters a clear description of the case to search for similiar cases.
The fact part of case documents in LeCard \cite{LeCard}, i.e., the \textbf{case description}, has an average length of 6333.7 words as shown in Table~\ref{QWLengthCaseDescriptionLength}. 
Therefore, both the queries and candidate results in legal case retrieval tend to be extremely long in practice than traditional Web retrieval. 



\section{proposed framework}

Legal case retrieval differs from Web retrieval in many aspects including query words, targets, and professional degree of users\cite{LegalSearchBehaviours1}. 

We conducted statistics on the distribution of the length of the judgments and queries in LeCard. These judgments will be provided to users as search results~(like candidate documents in web search). The results are shown in Table \ref{QWLengthCaseDescriptionLength}. 

\begin{table}[ht]
    \caption{Statistical quantities of the \textit{length of queries} and \textit{length of judgments} in the LeCard dataset. (expressed in words)}
    \centering
    \begin{tabular}{lrr}
    \toprule
& Query& Judgement\\
    \midrule
        Average Length &$6,333$ &$8,274$\\
        Min Length &$58$&$541$\\
        Median Length &$3,163$ &$4,804$\\
        Max Length &$90,181$&$99,163$\\
    \bottomrule
    \end{tabular}
    \label{QWLengthCaseDescriptionLength}
\end{table}

Methods of using traditional signals such as clicks and eye movements for relevance feedback in traditional web retrieval is difficult to apply in legal case retrieval for two reasons. 
(1)~In legal case retrieval, search results are relatively long(as shown in Table \ref{QWLengthCaseDescriptionLength}), hence searchers may need to click every search results and examine for a long time before they make a judgment.
(2)~Users do not behave the same way when reading this text, which has a fixed format and is thousands of words long~(details in experiments), as they do when reading a normal web page~\cite{InvestigatingUserBehaviorinLegalCaseRetrievalSYQ,LegalSearchBehaviours2}.

\begin{figure}[ht]
    \centering
    \includegraphics[scale=0.56]{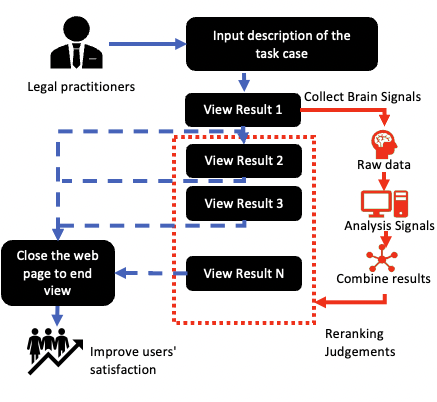}
    \caption{Legal case retrieval system supported by brain signals. The user inputs the description of her task case to begin the search process. \textcolor{blue}{Blue} parts show the user's procedure when using this framework. \textcolor{red}{Red} parts shows that during her judgments, EEG signals are collected and analyzed by the system. Based on the the analysis result, the system provides re-ranked results to the user.}
    \label{fig:system}
\end{figure}

Therefore, we propose a novel legal case retrieval framework with support of EEG signal, as shown in Figure \ref{fig:system}. When using this system, the user needs to wear an EEG cap additionally, which allows the system to capture users' EEG signals.

During the search process, we do NOT place restrictions on the user's behavior. Once the professional user enters the query, the system shows the most useful judgment among all retrieved. As the user reads this result, the system captures and analyzes her EEG signal. This process usually lasts for several minutes. After the user has viewed the first judgment, he can choose to continue to view the next judgment for more information. He can also choose not to continue browsing and simply close the web page.

EEG signals are continuously recorded throughout the user's browsing of the class case. After the user finishes a "paragraph" view, we store and analyze the EEG signals during this period as a separate segment of data. The EEG signal of each paragraph will be stored and analysed separately. The EEG analysis module will analyze this segment of the signal and predict whether the user is satisfied with this segment. 

The EEG analysis module in the system then monitor this process. The module does not interrupt users from making judgments. Instead, the module analyzes the user's EEG signals and determines that the user is satisfied with the first judgment or not. Afterwards, the system re-rank other judgments that have not yet been displayed based on the results of the analysis. 



\begin{figure}[ht]
    \centering
    \includegraphics[scale=0.24]{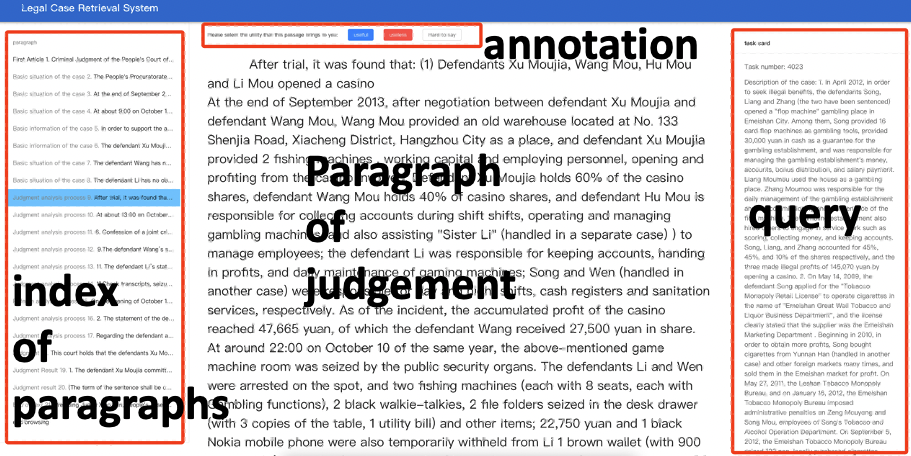}
    \caption{Example of the page in which users view the judgment~(translated).}
    \label{DocPage}
\end{figure}

Also, to easily record the EEG signals of participants, the following settings were made in the experimental system.
Results in Table \ref{QWLengthCaseDescriptionLength} denote that a judgement cannot be displayed in a single screen~(because of its length). When reading a judgement, the user may need to read it by paragraph. It is necessary to provide an index by paragraph reader for users. Table \ref{QWLengthCaseDescriptionLength} also shows that when using case descriptions as input, the length is much longer than the ordinary query words for web search. Since the query is long, we display it on the result page so that the user can compare the case with the result judgement.
We divide each judgement into several paragraphs. The standard for dividing paragraphs is that each paragraph is no more than 500 words (approximately one screen) and that the same sentence will not be divided into two paragraphs. This makes it easy for the system to record the EEG signals of users as they read each paragraph. 
A simple demo is shown in Figure \ref{DocPage}. In index, the paragraphs id and preview are displayed.

%% file: 5methods.tex
\section{Feedback Prediction with EEG}
\subsection{Limitations of Traditional EEG Methods}
\begin{figure}[ht]
    \hspace*{\fill}%
    \subfloat[Traditional ERP.\label{fig:erp1}]
    {\includegraphics[width=.502\linewidth]{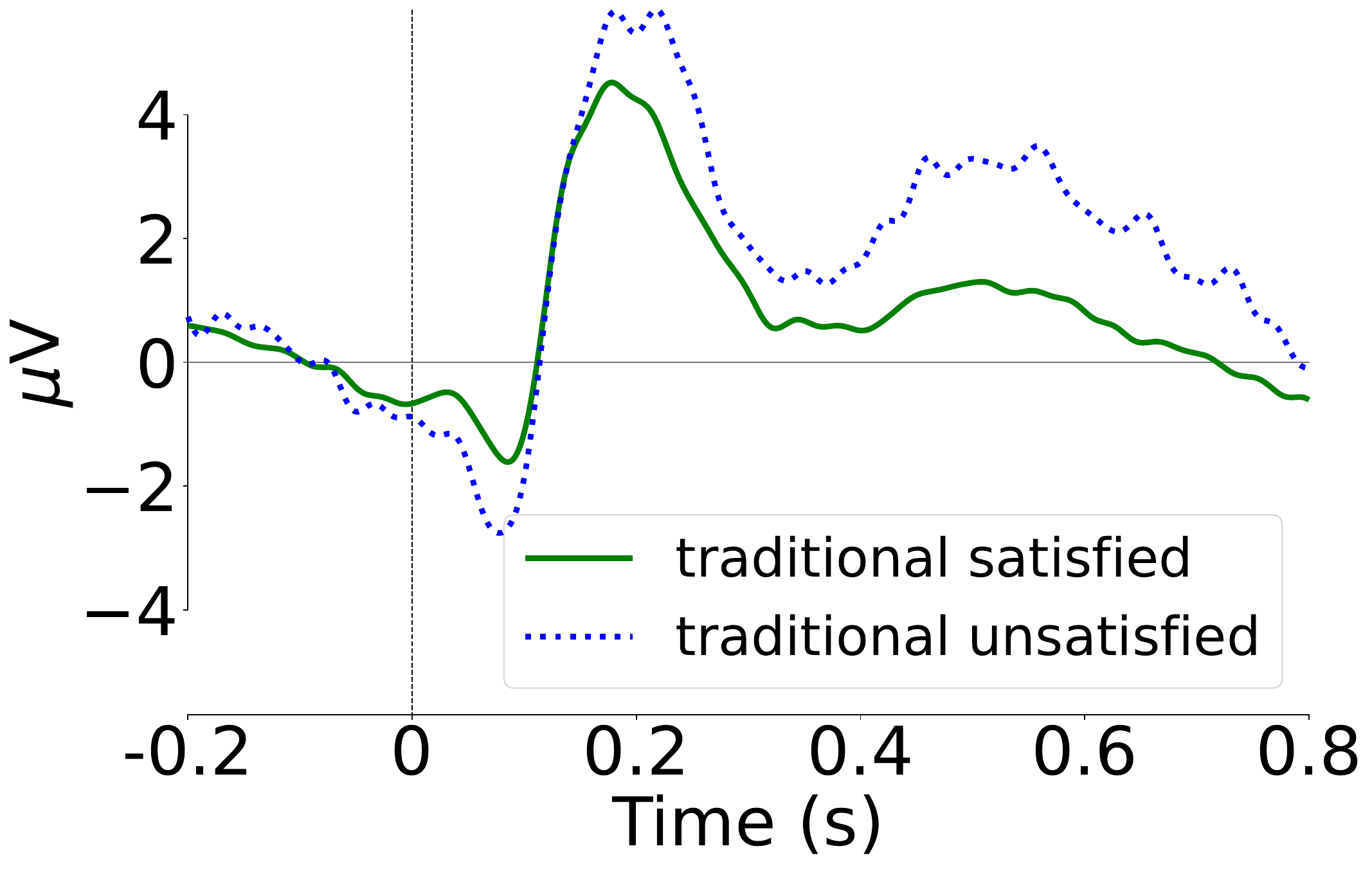}}
    \hfill
    \subfloat[Legal ERP.\label{fig:erp2}]
    {\includegraphics[width=.488\linewidth]{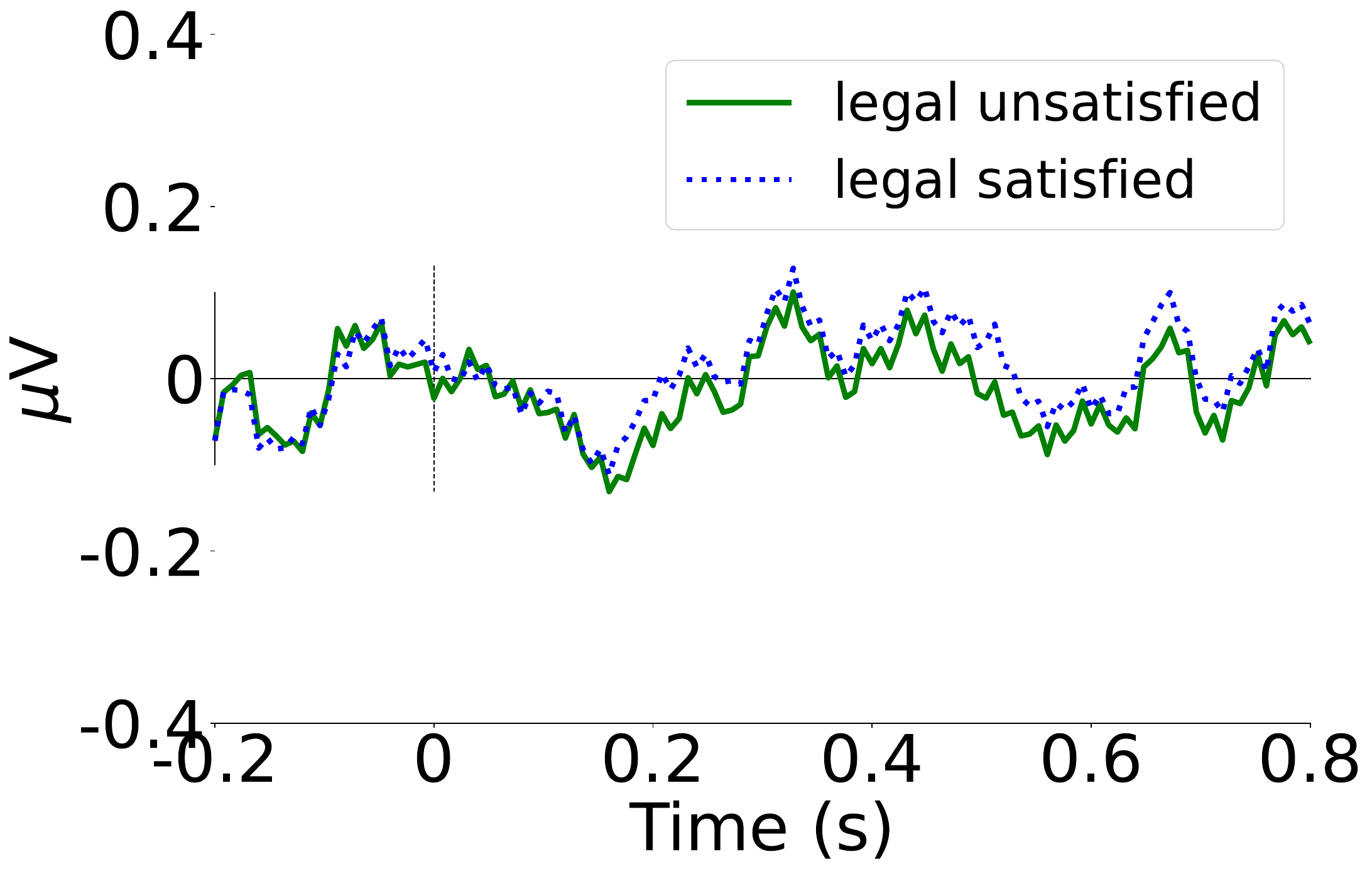}}
    \vspace{-3mm}
    \caption{Comparison of Event-Related-Potential~(ERP) in Web search scenario and ERP in leagl search scenario.}
   	\label{fig:erp}
\end{figure}

In traditional EEG methods, scientists usually record the EEG signal within $0.5$ to $3$ seconds of stimulation in order to analyze it.
This method is known as Event-Related-Potential~(ERP) analyses~\cite{sur2009event}, which analyze the EEG signals upon the stimuli presented and satcked data samples with the same type of stimuli~(e.g., unstaisfied or satisfied).
For example, Figure~\ref{fig:erp} presents two typical examples of ERP analyses in general IR domain~(the ERP data is inherited from \citet{ye2022towards}, as shown in Figure~\ref{fig:erp1}) and in legal domain~(as shown in Figure~\ref{fig:erp2}).
As shown in Figure~\ref{fig:erp1}, we can observe significant differences between traditional unsatisfied and satisfied ERP samples, especially in ERP components of 100ms, 200ms, and time points after 400ms.
The observation regarding the presentation time is known as ERP components, which can be applied to distinguish unsatisfied and satisfied data samples.
The ERP approach has been prevalently applied to study the phenomenon of relevance judgment and it has been observed that relevance judgment evolves time windows from around 200ms to 800ms~\cite{ye2022towards, allegretti2015relevance}.
However, this approach doesn't work among users of the legal case retrieval. 
As shown in Figure~\ref{fig:erp2}, similar ERP peaks may not exist in the data collected from participants during legal case retrieval.
As shown in Figure~\ref{fig:erp2}, although we have stacked the ERP data among thousands of data samples, the ERP in legal scenarios are still similar among brain responses to satisfied documents and unsatisfied documents.
Hence, we should design feature extraction methods differentiating from existing EEG methods.

\begin{table}
    \caption{Statistical quantities of the \textit{dwell time on judgements} and \textit{dwell time on paragraphs} in second(s).Q1 and Q3 are the first quartile(25\% percentile) and the third quartile(75\% percentile).}
    \centering
    \begin{tabular}{rrr}
    \toprule
& Paragraph& Judgement\\
    \midrule
        Average &$28.07$ &$311.21$\\
        Q1&$7.81$&$69.50$\\
        Median &$13.46$&$160.08$\\
        Q3&$26.68$&$401.668$\\
        
    \bottomrule
    \end{tabular}
    \label{dwelltimeParaDoc}
\end{table}

Further, we statistics the dwell time on a paragraph and a judgment of legal case retrieval users in the experiment.
The results, shown in Table \ref{dwelltimeParaDoc}, clearly show that users stay on a judgment for about 5 minutes in average, which is much longer than the time they spend on a single web page.
This suggest that extracting features from a longer time duration than previous research is another challenging problem.

\subsection{Feature Extraction}
 We propose a novel method for extracting features from the EEG data of users who retrieve legal cases.
 As shown above, traditional methods for feature extraction of EEG signals are not feasible in legal case retrieval. 
 Therefore, it is necessary to develop a novel algorithm that can effectively extract the features of EEG signals in legal case retrieval.

The algorithm we propose is based on the following two main ideas to extract features. 

First, participants were browsing judgments for a rather long period of time. During this time , participants were not always gaining useful information. From the interviews with the participants, we learned that for the majority of the time, all participants were "\textbf{looking for}" information that might help them solve the problem.
This means that it was unknown to the participants whether they could find useful information during this time. 
Hence, the stimulus including useful information only appears in a short period in the long period of reading time in legal case retrieval.

Previous studies \cite{moshfeghi2016understanding,paisalnan2022neural,pinkosova2020cortical,gwizdka2017temporal}  have shown that effective EEG features can be extracted if the stimulus happens in a short time period. 
Specifically, they used the following paradigm to extract features from a short-time stimulus:
\begin{itemize}
    \item Intercept the EEG signal for a short period of time~($1$ to $3$ seconds) after the subject is stimulated.
    \item This small period of EEG signal is pre-processed to remove the effects of artifacts such as motor movements, environment noises.
    \item The features of EEG signals in the frequency domain were obtained by Fourier transform and divided into five bands($\delta=[0.5,4]Hz$,$\theta=[4,8]Hz$,$\alpha=[8,12]Hz$,$\beta=[12,30]Hz$,$\gamma=[30,45]Hz$). Then, the energy of the EEG signal under the five bands is calculated and averaged within each band.
\end{itemize}

Based on the two observations above, we propose to adapt existing EEG feature extraction methods to fit the need of long period EEG analysis in legal case retrieval. Considering that participants only acquired the useful information in a small part of the whole reading time, we first sliced the EEG signals of the participants during the whole reading time and searched for the desired signals from the snippets obtained by slicing.

Our feature extraction is divided into the following 3 steps, which we will introduce in the next 3 subsections in turn.
\begin{itemize}
    \item Split by sliding window.
    \item Calculate the energy matrix within a single sliding window.
    \item Combine the calculation results with statistic.
\end{itemize}
\subsubsection{Split by sliding window}
To simplify the formulation, we use the symbol $S$ to denote the brain signal(EEG) of the user $u$ while he/she reading the candidate judgement with index $i$ with the query’s index is $j$. 
Here, the EEG signal S is a digital matrix of $ch \times (sr\cdot secs)$ given by the EEG acquisition device. 
Where $ch$ is the number of channels which is $62$ in our experiment.(There are $64$ electrodes (channels) on our EEG cap, including $1$ ground electrode, $1$ reference electrode and $62$ electrodes for recording data.) $sr$ is the sample rate of the EEG device, i.e., how many times the EEG signal will be sampled by the EEG device per second. 
This value should be set to a frequency higher than $45\times 2=90 Hz$ by Nyquist–Shannon sampling theorem \cite{Shannon1949}. 
$secs$ is the is the length of time (in seconds) that the participant stays on the judgement. 
The elements $v_{l,r}$ of the $l-th$ row and $r-th$ column of the matrix are the voltage signals on the participant's surface of the head acquired by the $l-th$ channel of the EEG device at moment $r$.

\begin{align}
    S=s_{u,i,j}=
    \begin{bmatrix}
        v_{1,1}&\cdots&v_{1,(sr\cdot secs)}\\
        \vdots& &\vdots\\
        v_{ch,1}&\cdots&v_{ch,(sr\cdot secs)}\\
    \end{bmatrix}_{ch\times (sr\cdot secs)}
\end{align}

Considering that in the legal case retrieval task, participants tend to stay on a single paragraph for a longer time than Web retrieval, we performed a "sliding window splitting" of this EEG matrix.
In Table~\ref{dwelltimeParaDoc}, we show the time that participants stayed within a single paragraph based on the dataset obtained from the user experiment. The statistical results show that the time that users stay within a single paragraph is much longer than the 0.5-5 seconds analyzed by the previous EEG model. Therefore the whole matrix needs to be sliced into sub-matrices of appropriate size in order to analyze.

The energy of the EEG signal over a continuous period of time often reflects the participants' logical thinking during that period of time. 
We want to capture the features reflected by participants' EEG signals in each small time slice, but we do not want to cut off the useful continuous segments from the middle because of our slicing. 
Therefore, we used a sliding window pattern to divide $S$. 
Specifically, consistent with previous work, we first intercepted the EEG signal within $t$ seconds($t$ is a hyperparameter, which in our experiments was chosen to be $1/2/4/8$) after the user started reading the segment as the first submatrix. 
Then, we slide the window by $1$ second to intercept the second submatrix, i.e., the EEG signal at the $[1,t+1]$ second. 
We continue this process by intercepting the submatrix at time $[2,t+2],[3,t+3]...[k,t+k]$ until the window can't slide further back (i.e. $t+k\leq secs<t+k+1$). Formally, we can get $k+1$ matrices of the same size $ch\times (sr\cdot t)$, which we call $S_i(0\leq i\leq k)$.
\begin{align}
    S_i=
    \begin{bmatrix}
        v_{1,(sr\cdot i + 1)}&\cdots&v_{1,(sr\cdot (i+t))}\\
        \vdots& &\vdots\\
        v_{ch,(sr\cdot i + 1)}&\cdots&v_{ch,(sr\cdot (i+t))}\\
    \end{bmatrix}_{ch\times (sr\cdot t)}
\end{align}

\subsubsection{Calculate the energy matrix within a single sliding window}
Next, we calculate the energy matrix $E_i$ for each of submatrices $S_i$. In matrix $S_i$, the data in the $r$-th row represent the voltage at each instant of channel $r$ during this period of time. For each channel's voltage sequence, we obtain its spectral sequence by discrete Fourier transform~\cite{DFT}. We do this for each row, and we get a spectral matrix $F_i$. Expressed in the formula~\ref{DFT-mat}. For simplicity, we denote the $r$-th row and $c$-th column of the result matrix as $f_{r,c}$.

\begin{align}
    \label{DFT-mat}
    F_i&=
    \begin{bmatrix}
        f_{1,1}&\cdots & f_{1,(sr\cdot t)}\\
        \vdots& &\vdots\\
        f_{ch,1}&\cdots & f_{ch,(sr\cdot t)}\\
    \end{bmatrix}_{ch\times (sr\cdot t)}\nonumber\\
    &=DFT(S_i)\nonumber\\
    &=
    \begin{bmatrix}
        DFT(v_{1,[(sr\cdot i + 1),(sr\cdot (i+t))]})\\
        \vdots\\
        DFT(v_{ch,[(sr\cdot i + 1),(sr\cdot (i+t))]})\\
    \end{bmatrix}_{ch\times (sr\cdot t)}
\end{align}

Where $DFT(*)$ denotes the discrete Fourier transform, the input is a time-domain signal sequence, and the output is a spectral sequence of equal length. 
In the spectrum sequence, the $i$-th($1< i \leq sr\cdot t$) value represents the intensity of that signal at $(i-1) Hz$ and, in particular, the first value represents the DC bias of that signal ($0$ in the data after proper preprocessing). The unit of it is also $Volt(V)$. 
Next, we calculate the energy density matrix $P_i$ from the $F_i$ matrix.
Square each element of the $F$ matrix to obtain the energy density matrix $P_i$, as described in Equation \ref{energy}.
\begin{align}
    \label{energy}
    P_i=F_i^{\text{\textcircled{\begin{small}2\end{small}}}}=
    \begin{bmatrix}
        f_{1,1}^2&\cdots & f_{1,(sr\cdot t)}^2\\
        \vdots& &\vdots\\
        f_{ch,1}^2&\cdots & f_{ch,(sr\cdot t)}^2\\
    \end{bmatrix}_{ch\times (sr\cdot t)}
\end{align}
For the energy density matrix $P_i$, the elements of the $r$-th row and $c$-th column($1\leq r\leq ch \wedge 2\leq c\leq (sr*t)$) represent the energy of the $r$-th channel at frequency $c-1 Hz$ in $V^2$.

Next, we just need to add up the values in the energy matrix $P_i$ in the corresponding interval to get the energy in the corresponding EEG frequency interval. According to five bands of EEG($\delta=[0.5,4]Hz$,$\theta=[4,8]Hz$,$\alpha=[8,12]Hz$,$\beta=[12,30]Hz$,$\gamma=[30,45]Hz$), we sum the values in the five intervals ($[2,5]$, $[5,9]$, $[9,13]$, $[13,31]$ and $[31,46]$) in each row to obtain five values that represent the sum of the energies in the interval. Here, we get the energy matrix \ref{Ei} which we need with 5 columns.

\begin{align}
    \label{Ei}
    E_i=
    \begin{bmatrix}
        \delta_{1}&\theta_{1}&\alpha_{1}&\beta_{1}&\gamma_{1}\\
        \delta_{2}&\theta_{2}&\alpha_{2}&\beta_{2}&\gamma_{2}\\
        &&\vdots&&\\
        \delta_{ch}&\theta_{ch}&\alpha_{ch}&\beta_{ch}&\gamma_{ch}\\
    \end{bmatrix}_{ch\times 5}
\end{align}

\subsubsection{Combine the calculation results with statistic}

Next, we merge results from the previous step to obtain the feature set of the whole EEG signal.
In the previous step, we extracted the feature matrix $E_i(1\leq i \leq k)$ for each slice of time. in this step, we need to merge them in order to extract useful features and reduce unnecessary redundancy.

Here, we consider these three challenges.

\textbf{Sparsity} First, consider that the user only gets information in a short period of time($5\%$or less of total time). The energy of these signals may deviate more from the mean value, and our algorithm needs to be able to keep these signals.

\textbf{Uncertainty} Second, on interviews with users, we notice that user do not have a certain time point to get useful information. They may be at the beginning of the reading, in the middle of the reading or near the end of the reading.Therefore, we would like to enable the algorithm to extract the feature signals that occur at different moments.

\textbf{Noiseability} Third, it is also important to note that the EEG device collects some noise signals when it acquires the EEG signals. These noises may come from the environment, EMG, or ECG. Although the noise is being filtered as much as possible during pre-processing, there is inevitably some residual noise. Therefore, it is necessary to consider how to reduce the effect caused by noise as well.

We use the following design to deal with these challenges.

The size of all matrices $E_i(1\leq i\leq k)$ is $ch*5$. The elements of the $r$-th row and $c$-th column of the matrix express the energy of the $r$-th channel at the $c$-th spectrum range in time slice $i$. We use the variable $E_i[r,c]$ to denote it.
Here, we use the formula \ref{comb} to extract features.Where $g$ is a set of hyperparameters ($[1,2,4,8]$ in our experiments), $max(E_*[r,c],g_j)$ means to take the $g_j$-th largest value in $E_i[r,c]~(1\leq i\leq k)$, and $min(E_*[r,c],g_j)$ means to take the $g_j$-th smallest value in $E_i[r,c]~(1\leq i\leq k)$.(If there is not enough $E_i$($k<g_j$), then we set $G_{max}[r,c,j]=G_{min}[r,c,j]=0$.) Finally, we get $2|g|$ parameters $G_{max}[r,c,j],G_{min}[r,c,j]~(1\leq j\leq |g|)$.

\begin{align}
    \label{comb}
    G_{max}[r,c,j]=max(E_*[r,c],g_j)&\nonumber\\
    G_{min}[r,c,j]=min(E_*[r,c],g_j)\}&
    (1\leq j\leq |g|)
\end{align}

This combination method is a good solution for these 3 challenges.
First, the statistical method of taking the kth value allows the \textbf{not for long} EEG signal to be captured. Second, the $k$-th value is taken from all $E_i$ sets when extracting features, independent of the order, which solves the \textbf{No confirmed time} problem. Third, the hyperparameter $g$ can be changed to further reduce the effect of \textbf{Noise} (e.g., when $g_j=2$, the extreme values of a signal point caused by noise do not affect the feature extractions).

It is worth to note that the size t of the slices also has influence on the list of extracted features results. Therefore, we set a hyperparameter list $T$($[1,2,4,8]$ in our experiment), which allows us to extract features of different size by setting $t=T_i$, and then merge features from diffrent $t$.

By modifying the hyperparameters, we can obtain a rather large list of features. The size of the feature list is $2\cdot |g| \cdot |T| \cdot ch \cdot 5$.Where $g,ch,T$ are defined as described above. For example, as the setting in our experiments, $g=\{1,2,4,8\},T=\{1,2,4,8\},ch=62$. Then the number of features are $2\cdot 4\cdot 4\cdot 62\cdot 5=9920$ in total.



The relevance feedback prediction is in two steps. In the first step, we use features to predict the relevance feedback for each paragraph. In the second step, we use "voting" to get the relevance feedback of the full judgement.

First, we describe how we use these features to train the model and finally make predictions on the usefulness of paragraphs.In the previous section, we extracted a large number of features from the EEG signal. However, we know nothing about which features are really useful. There is certainly useless and redundant features. We try to solve this problem by using machine learning models.

Here, we choose SVM-RFE~(support vector machine recursive feature elimination)~\cite{svm-rfe} to make predictions whether the paragraph is useful to current user or not. SVM-RFE is a supervised machine learning method based on SVM. In each round of iterations, SVM-RFE first fitting results using SVM, and then removing several features with lowest weights in the fitting function.

Next, we use a threshold $threshold$ to predict the satisfaction of the whole judgment. Users are satisfied with the whole judgment if and only if the number of satisfied paragraphs is greater than or equal to $threshold$. We set $threshold=3$ in our experiments. More about $threshold$ will be show in experiment.


\section{Topic-based Relevance Feedback}

The SVM-RFE predicts whether the user is satisfied with her previously read judgment. Based on this relevance feedback predicted by EEG signals, we adjust the display order of the remaining results.

In this paper, we use the labels in LeCard \cite{LeCard} and the labels added by \citet{zhang2023diverse}. LeCard is a legal case retrieval dataset which contains 107 queries, each of them is a case description. For each query, 30 candidate judgments are provided. LeCard provides a relevance label (denoted by 1 to 4) between each query-judgment pair. This relevance is generalized and does not consider the user's intent. In the \citet{zhang2023diverse}'s work based on LeCard, intents of each query are labeled. 
They use crimes as subtopics of judgment. They use $I_i\in [0,1]$ to denote the intent level of the i-th intent of the query. And  $D_{j,i}\in [0,1]$ are used to denote the j-th judgement's relevance for the i-th intent. 

In this paper, we follow this setting and score candidate judgments with these two labels. We suggest that the ranking score of a judgment can be expressed as the summary of all its relevance for each intent times intent level, as show in align~\ref{summ}.

\begin{align}
    RankingScore_j = \sum_{i \in intent} I_i\cdot D_{j,i}
    \label{summ}
\end{align}

The judgment with the highest ranking score was selected as the first judgment for users to read.
After the user has read a judgment, if the system predicts that the user is satisfied, it keeps all the values of $I_i$ unchanged. 
Otherwise, we consider that the user is not satisfied with the current judgment and attribute it to the top-t intent with the highest weight (in our experiments, we set $t=1$) and halve the value of $I_i$ for these intents. 
Then, the ranking score of all rest judgments will be recalculated based on the adjusted $I_i$. 


%% file: 6dataCollection.tex
\section{data collection}

To the best of our knowledge, there is no publicly available EEG dataset collected in legal case retrieval scenarios. Hence we propose this innovation work. We conducted numerous user experiments and collected, structured, and made public a EEG dataset.

\subsection{Tasks}

We selected 4 queries from LeCard dataset as the pre-defined task for this experiment. Then, we rewrote the case description of original queries, deleted parts that were not relevant to legal case retrieval, and reduced the length of the text to help participants understand the task more quickly. 
One of these tasks was used to guide participants to adapt to the environment of the experiment system and is always presented first. 
Three other tasks were used as the official experimental tasks. 
For details of these 3 tasks see Table~\ref{tasktable}. 
In all tasks, there is a pre-defined case description describing the facts of a complex case. 
For each task, seven judgments were selected as candidate judgments for our experiments according to the following rules. 
We first selected the judgement with the maximum $d_j,i$ for each intent. Then, we selected most relevance judgement by label $r_i$ in LeCard. 
If the user's intention is focus on any crime~(subtopic) which is related to the query, the list of candidate judgements will have at least one results related to that crime as well.

\begin{table*}
    \caption{Legal cases of experiments. TaskId is a unique id in Lecard. According to LeCard, we use crimes as subtopics (intents of the task). Origin length is the length of case description in LeCard. `Simple Length` is the length after rewriting.}
    \centering
    \begin{tabular}{cccccc}
        \toprule
        TaskId & &Crimes& &Revised Length& Original Length \\
        \midrule
        4023 & Running gambling house& Illegal business operation&-& 576& 751\\
        6432 &Bribery &Abuse of power & -& 501& 1195\\
        4863 &Picking quarrels and provoking trouble &Intentional destroying property & Intentional injury& 241& 280\\
        \bottomrule
    \end{tabular}
    \label{tasktable}
\end{table*}
\subsection{Participants}
We recruited 20 participants~(10males, 10females) via online forms and social networks. All 20 participants were graduate students in law school and aged 18 to 24(22.6 in average). They are all native Chinese speakers. Moreover, we required that participants already be licensed to practice law~(passed "National Uniform Legal Profession Qualification Examination") or have the same level of knowledge.

We divided participants into 2 groups. 
One group of participants reads the judgments in a fixed order from high to low relevance.
In the other group, after the first judgement is read, the remaining 6 candidate judgements will be re-ranked according to the method in Section 4. 
Unfortunately, because we couldn't build the EEG analysis beforehand, we have to use the relevance feedback collected from the participant's explicit annotations for result re-ranking in the user study. 
However, after we build the EEG module, we can use it to predict feedback results and evaluate it with a simulated experiment constructed with the user study data. More details can be found in Section 6. 

\subsection{Procedure}

Before the experiment starts, participants were given an EEG cap that fit their head size and connected to the EEG acquisition system. Then, participants were asked to read an instruction that outlined the specific tasks to be performed and informed them of the details of tasks.
Participants were asked to assume that they became the lawyer or judge of the case and needed to simulate the behavior of a real lawyer or judge when doing legal case retrieval tasks.

\begin{figure}[ht]
    \includegraphics[scale=0.25]{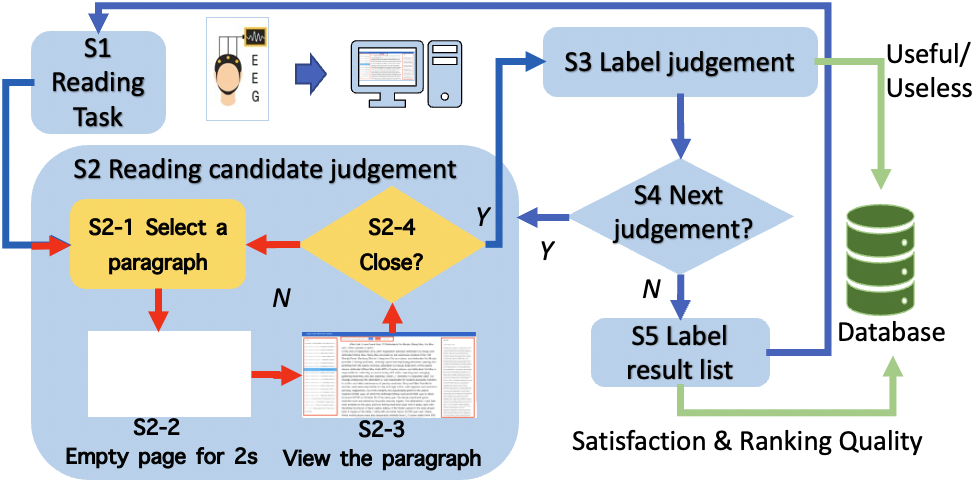}
    \caption{Experimental process. \textcolor{blue}{Blue} lines shows the process of the participants' interactions. \textcolor{red}{Red} lines shows the sub-process in S2. \textcolor{green}{Green} lines are the data flow of annotations. S2 stands for a loop of reading paragraphs of single judgement, while S2-S3-S4 stands for a large loop of reading multiple judgements of a single task. The process of S1-S2-S3-S4-S5 stands for a tasks' process and will repeat 4 times for each participant. First, participants read the task description~(S1). Second, participants read the first judgment given by the system~(S2). During the reading process, subjects could choose to read one of the paragraphs each time. And they are allowed to end the reading if they think enough information is acquired. Third, participants annotated the judgment they had just read~(S3). Fourth, participants choose whether to read next judgment~(S4). Fifth, participants mark the completion of the whole task~(S5).}
    \label{ExpProduce}
\end{figure}
During the experiment, the procedures of participants are shown in Figure \ref{ExpProduce}. The EEG signal of participants will be recorded in the whole process. Then, we introduce the details of each part of the experiment. 

In Step1 (S1 in Figure), the participant is asked to read the task description that we have prepared previously. In Step1 (S1 in Figure), the participant is asked to read the task description that we have prepared previously. We asked participants to think about which crimes this case description might be related to during this process. Participants were required to read  judgments and analyze the task in a following process. Task description is shortened from LeCard's query. After reading, the participant clicked the "Next" button to go to Step2.

The Step2 is a cyclic process in which the participant reads one judgment. 
Specifically, the page was organized as shown in Figure \ref{DocPage}. 
Subjects clicked on any of the passages in the index on the left side to read one paragraph each time. Participants can read any paragraphs in any order. The " Close " button is located in the left bottom corner of the page.

In Step3, the participant is asked to evaluate each of the paragraphs she has read in Step2. Three options available are: useful for analysis tasks, useless for analysis tasks, hard to say.

In Step4, the participant is asked whether to continue reading another judgment. Participants could choose NOT(ending the reading of this task) in two situations. \textbf{A}. The previously read judgments were enough in analyzing the case in the task. \textbf{B}. Further reading would not provide further assistance in analyzing the case in the task.

In Step5, the participant is asked to label the following two contents. 1. In step 4, for what reason did you choose not to continue reading, A or B. 2. whether the order in which the candidate judgments were presented was well arranged. A 4-level labeling approach is used here.

Finally, at the end of Step5, we ask the participant to present a brief analysis of the task. The participant orally describes this content and is recorded, and the recording files are organized as a part of the data set.

The length of the recording for each task was 15-45 minutes. Afterwards, the participant takes a 3-minute break before next task.

After the experiment, we listened to the recordings of all participants to confirm that all participants completed the task seriously.

\subsection{Devices}
In this experiment, we have set up a desktop computer in the laboratory for participants to access the experimental system. A monitor with 27-inch with a resolution of 2560x1440 and a keyboard and a mouse were connected to this computer. (During the experiment, participants only use the mouse for operation, and the keyboard will not be used.) 

64-channel Quik-Cap (Compumedical NeuroScan) was worn by participants and used to capture EEG signals from participants. to A SynAmps RT \cite{eegcap}(Compumedics Ltd., VIC, Australia) is used to amplify signals from Quick-Cap, and output it with USB \cite{EEGAmplifier}. Software "Curry"(Compumedics Ltd.) is installed on the computer to collect data from USB port and store it as ``CNT'' format. The layout of the cap is according to the extended 10/20 system~\cite{homan1987cerebral}. The sample rate is set as 2000Hz. Before the experiment, we verify that the impedance of each electrode is less than 10 $k\Omega$. After the experiment, we repeat this to reconfirm that we have collected a valid data. 

\subsection{Preprocessing of EEG Data}
EEG data contains some noise, including line noise, blinks, muscle action, etc. We process raw data through a standard pre-processing procedure. The standard procedure \cite{UnderstandingNon-Click} included these in sequence: re-referencing  to averaged mastoids, baseline correlation, low-pass filtering at 50 Hz and high-pass filtering at 0.5 Hz,
removal of artifacts, down sampling to 1000Hz. Next, we segmented the data according to the time range in which participants viewed each paragraph. 

%% file: 7examinations.tex
\section{Experiments}

\subsection{Ex1:Predict Satisfaction on Paragraphs}

We designed an experiment to test the performance of models for predicting paragraph satisfaction based on EEG features.

We will use supervised learning methods to make predictions on paragraph satisfaction. We first divide the EEG signals of participants to obtain training set and test set. For each participant, we used the EEG signals when they performed the first two tasks as the training set and the EEG signals when they performed the last task as the test set. 
\begin{table}[ht]
    \caption{Accuracy of models in predicting paragraph satisfaction.}
    \centering
    \begin{tabular}{cccccc}
\toprule
 &Random& Linear & DecisionTree & MLP & SVM-RFE\\
\midrule
Acc. & 50\%&52\%&64\%&68\%&\textbf{71\%}\\
\bottomrule
    \end{tabular}
    \label{ex1}
\end{table}
In this experiment, the EEG signals of participants while reading each paragraph will be turned into a 10240-dimensional vectors using previous methods. 
We expect that the model we selected can be trained to predict the labels based on these vectors. 
The traning labels are the satisfaction of the paragraph given by participants.

We choose the following models for training and testing. All these models use code from scikit-learn~\cite{scikit-learn}.

\begin{itemize}
    \item \textbf{Linear Model} \cite{LinearReg} A linear regression approach to fitting labels.
    \item \textbf{Decision Trees} \cite{DecisionTree} A flowchart-like structure to classify samples step by step. We tried decision trees with depth 4,8,16,32 and selected the best of them with  depth 16.
    \item \textbf{SVM-RFE} \cite{svm-rfe} (Support Vector Machine Recursive Feature Elimination)~A support vector machine based feature extraction by iterating two steps of computing support vectors and removing the dimension with the lowest weights.
    \item \textbf{MLP} \cite{MLP} (Multilayer perceptron)~A neural network with multiple fully connected layers and activation functions.
\end{itemize}

The results of the experiment are shown in Table~\ref{ex1}. 
In our experiment, SVM-RFE is effective than other models, achieving an accuracy rate of 71\%. Decision trees and MLPs also achieved good results, while linear models performed poorly almost close to random.

\subsection{Ex2:Comparison to Relevance Feedback with Click and None Feedback}

We compared our feedback method among click-based feedback and none feedback.

We divided 20 participants into two groups of 10 participants each. 
One group of participants can acquire results that are re-ranked via relevance feedback with brain signals after participants read the first judgment, called \textbf{EEG}. 
The other group of participants is shown the results in a fixed ranking with documents from the highest relevance to the lowest relevance, called \textbf{None}.
And, we assume the third group. In this group, relevance feedback is predicted by the number of results that are clicked by users. A user is satisfied with the judgment if and only if the number of results clicked is greater or equal than $threshold$. We called this group \textbf{Click}. We set different thresholds to test these three methods. The prediction Accuracy and F1 Score are shown in Table \ref{ex3}. The results show that using EEG for relevance feedback is better than using clicks.


\begin{table}[ht]
    \caption{Performance of different relevance feedback methods. }
    \centering
    \begin{tabular}{cccc}
    \toprule
         & threshold&Acc.&F1 \\
         \midrule
        None &-&-&-\\
        \midrule
        Click &1&65\%&0.788\\
        Click &2&\textbf{70\%}&\textbf{0.813}\\
        Click &3&\textbf{70\%}&\textbf{0.813}\\
        \midrule
        EEG & 1&75\%&0.839\\
        EEG & 2&\textbf{90\%}&\textbf{0.929}\\
        EEG & 3&\textbf{90\%}&0.923\\
        \bottomrule
    \end{tabular}
    \label{ex3}
\end{table}
\begin{table}[ht]
    \caption{Effect of re-ranking on ranking quality evaluation and users' satisfaction.$*$ indicates $p<0.01$ and $**$ indicates $p<0.005$ compared to None.}
    \centering
    \begin{tabular}{lll}
        \toprule
        &Ranking Quality & Satisfied\\
        \midrule
        None & $2.53$&$43\%$ \\
        EEG &$2.80^*$&$80\%^{**}$\\
        \bottomrule
    \end{tabular}
    \label{ex2}
\end{table}

Moreover, labels of re-ranking given by participants were also collected, includes \textbf{Ranking Quality} and \textbf{Satisfied}. Ranking Quality is rated in 4-level(1 to 4 from "bad" to "good") to describe the display order facilitates for the task. And Satisfied is rated as satisfied(1) or unsatisfied(0) to describe if the results are helpful for the task. The results are shown in Table \ref{ex2}. It shows that, similar to web search, re-ranking based on relevance feedback in legal case search can improve user satisfaction significantly.

%% file: 8conclusion.tex
\section{conclusion and future work}
In summary, we propose a novel framework for legal case retrieval supported by brain signals. 
Further, we designed a series of user experiments based on this framework. 
We collated and made public the results of the user experiments and released the first EEG dataset for legal case retrieval.
Also, we designed an innovative feature extraction method, which extracted features with good performance in predicting users' relevance feedback.
We use two different re-ranking strategies in our user study, i.e., re-ranking with results diversity and results relevance.
If the participants marked the first results as irrelevant, than re-ranking with results diversity can provide diversify results to satisfy the user.
We verify that the re-ranking strategy can improve user satisfaction and simulated this process with brain signals is also helpful.

However, there are some limitations that can be further improved.
For example, developing feature extraction algorithms and relevance prediction models with better performance is a meaningful and challenging work.
New models may need to mining the useful features in the fine-grained EEG information of users on each paragraph to construct fine-grained retrieval models. And re-ranking methods also can be optimize.
Further, limited by the order of the experiment-model relationship, this work only simulated the environment for online prediction. It is also an exciting project to build an online, real-time legal case retrieval system and to collect more datasets.